\input epsf
\documentclass[nohyper,letterpaper,11pt]{JHEP}

\title{Light Front Models for the Leptons, Bosons, and Quarks}

\author{Roger C. Millikan\\
Department of Chemistry and Biochemistry\\
University of California Santa Barbara CA 93106-9510\\
millikan@chem.ucsb.edu}

\abstract{ The elementary particles are modeled as harmonic
oscillator excitations of transverse U(1) gauge fields propagating at
v = c, with open and closed string-like propagation paths. One, two
and three node states represent the leptons, bosons, and quarks. We
incorporate a twist  ($\theta$) for the gauge field components which
rotate counterclockwise (L) for the electron, yielding a chiral
model.  Theta increases by $\pi$ from node to node, making the lepton
 models SU(2) representations. At nodes the twist may reverse,
creating new particle states. For three nodes, twist combinations map
the SU(3) color states of the quarks.  Generations are modeled
topologically by the winding number of  the strings. Mapping model E
fields to distant observers makes  understandable how fractional
charges arise for the quarks. These models are 3D slices of
spacetime, allowing us to make drawings of particle field
conformations. From model particle  quantum numbers, new
mass relationships are derived.}

\preprint{Version 6/13/2001}
\begin{document}

\section{Introduction} 

The models of the elementary particles described here have drawn from
many antecedents. The original Skyrme model \cite{Skyr61} has been
developed in various directions, for example, by Thirring
\cite{Thir58} and Enz \cite{Enz77}. Those efforts inspire one to use
symmetries to generate boson and fermion particle states from
continuous spacetime fields in low dimensions. The recent and rapid
emergence of supersymmetric string models containing BPS states
protected from quantum corrections \cite{Sen95} encourages one to
believe that relatively simple models may indeed give good
representations of elementary particle states. From the quantum
gravity side, recent solutions of Einstein Yang Mills Dilaton field
equations \cite{Klei97} using axial symmetry have yielded particle
states with quantum numbers  and geometries similar to those of the
models presented here. And from Witten's \cite{Witt88}  development
of Chern-Simons topological field theory we have  learned that the
observable expectation values of relativistic quantum field theories
should be topological invariants. This view is reinforced by the
demand of relativists that theories of quantum gravity should be
diffeomorphism invariant \cite{Gamb96}. The real problem is how one
can pull all these threads together to weave a consistent theory that
represents nature as we find it. The standard model is a large step
in this direction. But its shortcomings and the need to move beyond
it have been well documented in the literature. 

Most current theoretical attacks start with postulated
actions or Lagrangians. One then struggles to see what particle
states and observables they produce. It now appears that the particle
states of interest cannot be reached by applying perturbation theory
to known, standard theories. Lacking that methodical approach, one
must be intuitive, clever or lucky in guessing new equations as
starting points to succeed in finding ``nature's equation". We
choose to reverse that approach, and begin by guessing the solutions
that account for known experimental results, and then work back
toward the equations. An unfailing guide in building such models is a
demand for consistency with all known experimental data, and with
cherished theoretical principles. The set of experimental data
constraining model building is much enlarged if one demands that all
known elementary particles (leptons, bosons, and gluon/quarks) must
be consistently described by one coherent model system. This is the
goal. 

We start with certain biases. Wishing to be ``physical", we confine
our modeling efforts to 3 + 1 dimensional spacetime and relevant
subspaces. We believe that topological and geometric features of
these spaces play a major role in determining the possible and
observable particle states. In keeping with this, we work in a
semi-classical mode. That is, while keeping in mind quantum demands
such as the importance of boundary conditions and wave function
nodes, we focus upon classical features dictated by the geometry and
topology of the models. Lastly, we believe it is important to be able
to visualize the models and their geometric features. To this end, we
have developed a depiction system for the models. 

Our light front model system for the elementary particles is
described below. We believe it succeeds on several levels. (1) It is
a consistent system that accommodates all the known elementary
particles in terms of a few quantum numbers. Those quantum numbers
have logical connections to the geometric, topological and harmonic
oscillator constraints that go into the theory. (2) It ties onto the
standard model and explicates how the U(1)  $\times$ SU(2) $\times$
SU(3) gauge groups arise. (3) It suggests answers to long standing
questions such as, ``What determines the differences between the
three generations of particles?". (4) It illuminates the origin of
electric charge, and makes the fractional charges of the quarks
understandable. (5) It provides a depiction system that makes it
possible to visualize each particle model. This makes it easy to
apprehend the relationships and properties of the different particle
states. (6) Studying the model system has suggested to us ways for
plotting the experimental rest masses against the model quantum
numbers for deriving the functional relationships. The parameters of
these mass plots only involve the natural constant $\alpha$ (or
e$^2$, the electroweak coupling constant)  and the electroweak mass
$m_W$.  

The plan of the paper is as follows: In section 2 we give the basic
assumptions that go into the model system. In section 3 the depiction
method is described and an overview of the particle models is
provided by a series of figures. Section 4 provides model details and
a discussion of salient features. The way the observed electric
charge for each particle arises is described. In section 5 we discuss
how the model is extended topologically to treat the higher
generations of fermion families. Section 6 summarizes the quantum
numbers that specify the particles. Mass relationships are developed
in section 7. In section 8 we consider some of the problems of
mapping from the spacetime of the particles to that of distant
observers. Section 9 presents our summary discussion. There follow
two Appendices. The first gives a \emph{Mathematica} notebook
providing calculational details to make explicit features of the
electron and quark models that are discussed only qualitatively in
the text. The second appendix gives explicit calculations of the
electric charges the models exhibit.

\section{Assumptions}

I. The most basic assumption we make is that
all observable particles are transverse quantum U(1) gauge
excitations moving at the local speed of light in 3 + 1 dimensional
spacetime. These excitations have two orthogonal vector field
components which may be identified as electric and magnetic fields in
the massive charged particles. In the large, the gauge excitations
propagate along geodesics of spacetime as demanded by general
relativity. In regions where the excitation is intense, the nonlinear
nature of the Yang Mills (and possibly dilaton) equations causes the
propagation path to contort and writhe, producing soliton
configurations. These are the particles. 

II. These excitations are best represented as quantum harmonic
oscillators having one node for the leptons, two nodes for the gauge
bosons, and three nodes for the quarks. Higher excitations may be
possible, but are presently unobserved.

III. The boundary conditions for the excitations are of two types. If
the wave functions fall asymptotically to zero at plus and minus
infinity (i.e. at a distant emission source and a distant absorber)
we have the zero rest mass particles (neutrinos, photons, and
gluons). If the wave functions are consistently self-bounded, we have
particles with finite rest mass (the electrons, the W and Z, and the
quarks). 

IV. The orthogonal components of the transverse gauge excitation have
a polarization degree of freedom denoted by the twist angle theta. In
the development of the models, we found that in every case theta had
to change by $\pi$ radians between nodes for the models to be
consistent. Hence we postulate that it is a rule of nature that 
theta must change by $\pi$ radians as the excitation wave function
advances from one node to the next. At a node, however, theta becomes
undefined. As the wave continues from a node, theta may twist as
before, or it may reverse. Which choice is made determines in large
measure which particle a given model represents. 

V. The non-linear nature of the equations causes the wave to contort
and spin about the geodesic, giving rise to spin angular momentum. 
The spinor particles with boundary condition $\psi (0) = -\psi(2\pi)$
have spin $\hbar/2$ and are fermions. The vector particles with boundary conditions
$\psi (0) = +\psi(2\pi)$ have spin $\hbar$ and are bosons.

\FIGURE{
\epsfbox{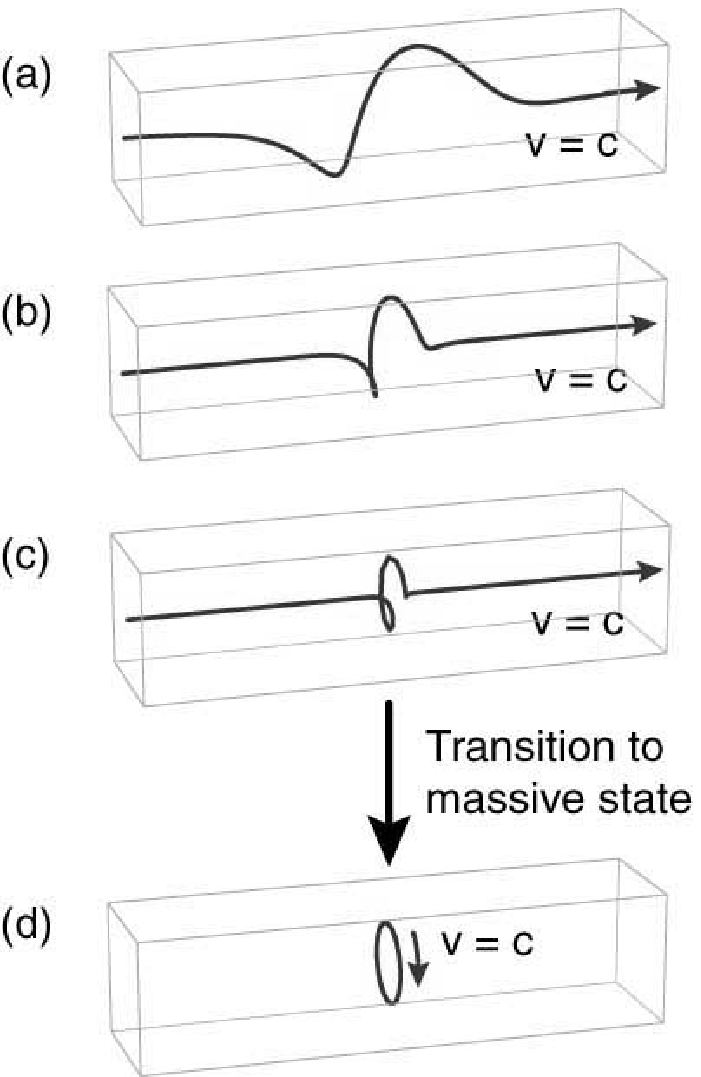}
\caption{ (a) Excitation with zero
rest mass propagating at v = c with energy E. (b) Energy = 3E. (c)
Energy = 9E. Axial extent of the excitation decreases with increasing
E but its curvature grows to keep the angular momentum constant at $n
\cdot \hbar$. (d) Excitation undergoes a transition to a self-bounded
topology with rest mass. (Propagation is still at v = c, but the
excitation remains in the volume element at rest relative to a
suitable massive observer.)}
}

VI. In higher energy particles, the propagation path of the
excitation may writhe to increase the winding number by an integer
number of units compared to the lowest energy particle of the same
field configuration. For winding numbers of 1, 2, and 3 we model the
observed family generations of fermions.

\section{Model Depiction and Overview} 
The known elementary particles
fall into two categories: those with zero rest mass that always
travel at v = c,  and those with finite rest mass  having v $<$ c.
Zero rest mass particles include the neutrinos, the photon and the
gluons. In the second category we find the massive leptons, the
massive gauge bosons, and the quarks. What distinguishes  these two
categories of particles? As stated in assumption III above, we
believe it is a question of boundary conditions for the wave
functions of the excitations. Consider a zero rest mass excitation
propagating at v = c. Freeze it at an instant of time on the light
cone. As depicted in Figure 1a, 1b, and 1c, its wave function must
die away smoothly and monotonically at large distances from the
location of the excitation. We use the local writhing of the wave
function to model the angular momentum of the excitation. Particles
have constant angular momentum n$\hbar$ independent of their energy.
This is incorporated in Fig. 1.

As the particle energy increases, the longitudinal extent of the
excitation decreases, but its curvature increases. Once the threshold
energy is passed where everything can fit, a topological transition
to a self-bounded state is possible as depicted in Fig. 1d. The
excitation now represents a particle with rest mass. The excitation
is still propagating at v = c, but since it is self-bounded, its
energy density stays in a small volume element of space relative to a
real external observer. This is what we mean by rest mass. Of course,
transitions to massive particles must conserve quantities like
electrical charge. Generally this means one must make a massive
particle and its antiparticle at the same time. To summarize, our
models of zero rest mass particles will utilize wave functions that
go smoothly to zero at plus and minus infinity along the propagation
path. Models of particles with rest mass will have wave functions
that close back on themselves and are self-bounded. In the standard
model, spontaneous symmetry breaking and the Higgs mechanism are used
to give mass to otherwise massless gauge particles. Perhaps this is
just a hidden way of introducing a topological boundary condition
change into the theory.\cite{Cast97}

The model fields to be depicted are moving at v = c. In order to see
them as static configurations, we imagine ourselves to be fiducial
observers moving at light velocity. That is, we place ourselves on a
slice of the light cone centered on the wave function of interest.
The x-axis of our image is oriented along the geodesic propagation
direction, and we assemble a volumetric image  by merging nearby
slices in the $+$x and $-$x directions. This is analogous to the way
Computer Aided Tomography (CATscan) images are assembled from a
series of cross sections. As the propagation path contorts, by
assumption I, the gauge field components must remain transverse. As a
result, we have two orthogonal components of a twisting field
geometry to depict. This is done using \emph{Mathematica} as a tool
for depicting 3D figures with hidden lines removed. (See Appendix 1).
On the model figures, it is useful to add an arrow which indicates
the direction of advance as the excitation propagates. For twisting
fields, this fixes the chirality.  

Our models of the massless particles use quantum harmonic oscillator
wave functions for both the amplitude of the field components and for
the radial displacement of the function baseline from the axial
propagation line. For the self-bounded models of the massive
particles, we use circular propagation paths and sine functions for
the amplitudes. The exact functions used are given in Appendix 1. In
the figures, we denote the negative half cycle of the E field vectors
using  a saturated orange color (dark gray shading for B\&W
reproduction). The positive E vectors are shown in light orange
(medium gray).  In a similar manner, for the N and S parts of the B
field we use saturated yellow and light yellow shading (light gray
and very light gray in B\&W reproduction).  

Using these  conventions, we show a collage of light front models of
selected elementary particles in Fig. 2.

\FIGURE{
\epsfbox{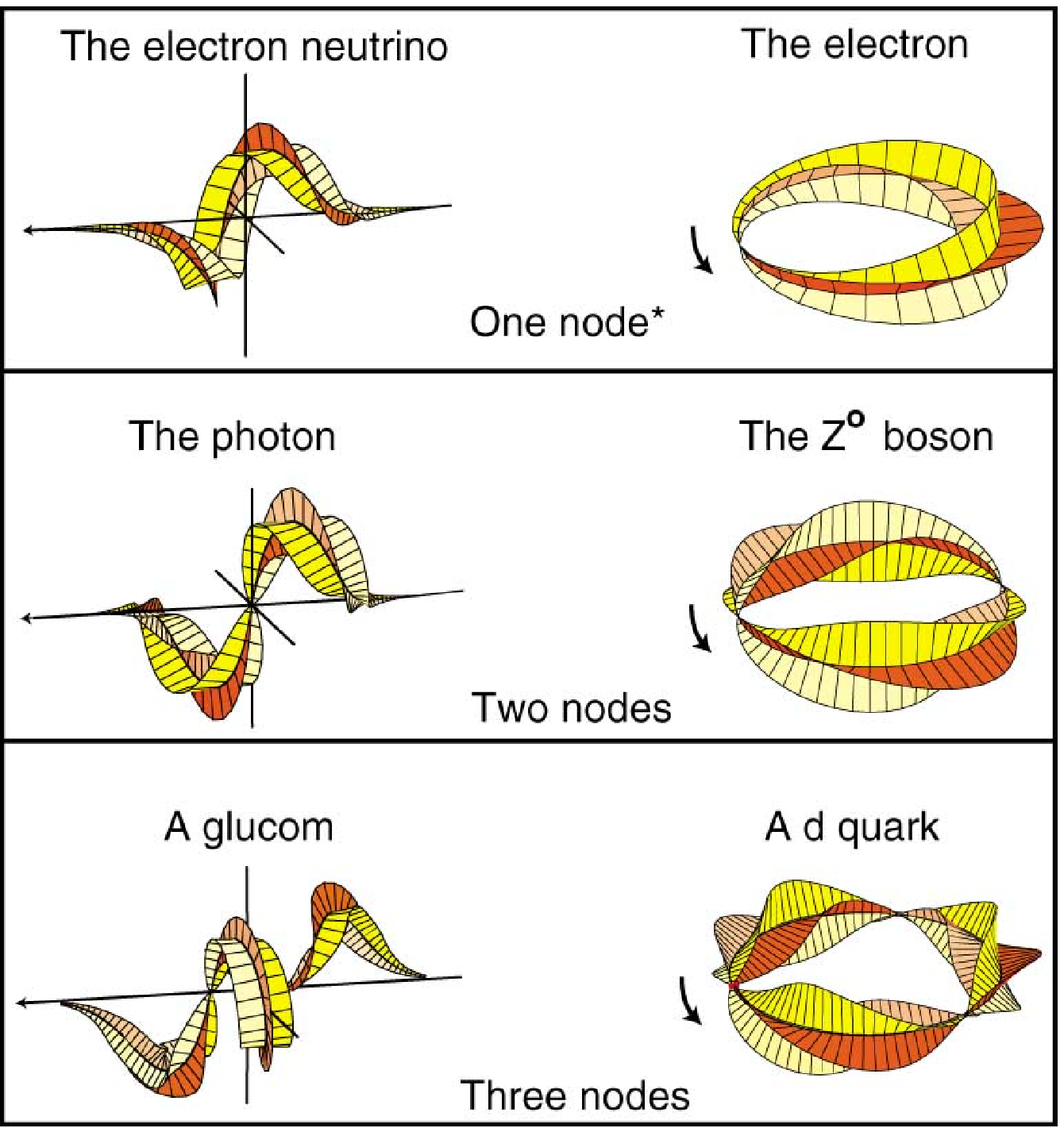}
\caption{A collage of light front models of
selected elementary particles. In the left column are zero rest mass
particles with wave functions bounded at plus and minus infinity. In
the right column are finite rest mass particles having self-bounded
wave functions. A glucom is a gluon component. These are first
generation particles. * We count one node at infinity.}
}

They are organized from top to bottom by the number of harmonic
oscillator nodes in the particle wave function.  The number of nodes
emerges as one of the key quantum numbers. Specifically, for
\textbf{n} = 1 we model the leptons, for \textbf{n} = 2 we model the
gauge bosons, and for \textbf{n} = 3 we model the quarks and glucoms
(which are gluon components to be discussed later). In keeping with
Assumption IV, the component field vectors twist about their
propagation path by $\pi$ radians as the wave front moves from a
given node to the next in the direction of the arrows. No
antiparticles are shown in Fig. 2, but they can easily be imagined.
To obtain the antiparticle of a given model one reverses the
direction of twist rotation as the fields advance. This is equivalent
to a PC transformation. It reverses the chirality and the charge
exhibited by the model.

In these models, there are three different features with periodicity.
The U(1) gauge fields are periodic. The twist angle theta is
periodic. And in the massive models, once around the circle is a
geometric period. The way these periodicities correlate differs from
particle model to particle model, and in fact determines the electric
charge exhibited. The story is simple but intriguing for the
\textbf{n} = 1 electron. As the number of nodes rises, the
possibilities increase, until we reach the quarks with their several
charges and multiple colors. Let us begin with the simple case first.

\section{Model details and salient features} 
\subsection {The electron model}
An oblique view of the electron model is shown at the upper
right in Fig. 2. The gauge fields are represented simply by
orthogonal sine waves propagating around a ring, twisting as they go.
The key feature is that the ring circumference is made equal to half
the electromagnetic wavelength. In similar manner, we make once
around the ring produce 1/2 twist, that is, $\pi$ radians. These are
considered quantum requirements which combine to produce a stable
soliton field configuration that models the electron. As a geometric
structure, it has some exceptional properties. The half twist for
once around makes each vector field into a M\"{o}bius strip. Normally,
such a manifold is non-orientable. But when the geometry inversion is
combined with the electromagnetic field reversal each time around,
the field becomes oriented and self-bounded. Consider the magnetic
field shown in yellow (light gray). Starting at the node, follow the 
wave function twisting around the ring as it goes through its maximum
and back to zero. Let this represent the north half cycle of the
field. Continuing around, the field reverses and we trace the light
yellow (lightest gray) wave function through its maximum and back to
zero. Let this represent the south half cycle of the field. At this
point we have completed one full cycle of magnetic field variation,
and also one full 2$\pi$ twist about the ring. So both the wave
function and the twist are ready to repeat their cyclic variation. In
other words, they are consistently self-bounded. Now look at the
resulting geometry. Even though the magnetic wave function is
alternating N to S in a cyclic way, the N half of the field is always
on the top side of the ring. In like manner, the S part of the field
is always on the bottom side of the ring. Recall that this is a light
front model with the fields propagating around the ring at v = c. From
far away, a real, massive observer would see only the time averaged
resultant of the field motion. He or she would see a magnetic dipole.
Now focus on the orthogonal fields shown in orange (dark gray). A
similar geometric effect occurs when the twist inversion is combined
with the cyclic field inversion. In this case one half of the field
variation (which we choose to call the negative half cycle since we
are modeling an electron) always remains external to the ring. The
other half cycle always remains internal to the ring. As before, we
start with an electromagnetic field component that is cyclically
varying from plus to minus, and end up with a time averaged field
that is minus external to the ring, and plus inside. What would a
distant real observer see? We claim he can see only the portion of
the field external to the ring, and consequently he sees a negatively
charged particle. The fields internal to the particle are hidden from
view. Why?  The ring is the locus where the fields are moving at the
local speed of light. This is very like a Kerr black hole which
possesses an event horizon. External observers cannot see any
phenomena that occur inside  such a horizon. We take this seriously,
and consider our model of the electron to be a soliton with an
extreme Kerr geometry. Such models (excluding the twist) have been
invoked before in the context of heterotic string solitons[4]. 

We note that once around the propagation ring corresponds to only
half the U(1) period. So the wave must go around again to make one
complete wavelength of the field. Requiring a 4$\pi$ revolution for a
symmetry operation is the hallmark of spinor behavior, and this
brings the fundamental representation of the SU(2) group into our
models. It is known from experiment that the electron neutrino and
the electron are chiral particles. Using standard conventions, they
are left handed. Our models have an internal chirality because of
their helical conformations. If we label them properly, they are left
handed. Further, as noted above, their antiparticle models are right
handed, in accord with experiment. 

We need to find a way to take the fields specified for the light front
models and deduce the magnitude of charge and dipole strength that
they represent to a real, distant observer. How does one measure the
charge contained in a volume of space? One integrates the field
strength emerging from a sphere containing the ``charge" and uses
Gauss' law to determine the number of charges inside. We will do
likewise for the models, using the following algorithm. The E field
vectors emerge from the ring at all angles, but we argue that only
the component of the field in the equatorial plane of the ring is
strictly radial, and capable of escaping the event horizon. So we
project each E vector onto the equatorial plane and integrate through
one complete cycle, keeping in the integral only those components
that are directed outwards. A visual way of doing this is to look
down onto the top view of the model, and count only the field
external to the ring. This calculation for the electron model of Fig.
2 yields the result of $\pi$, as detailed in Appendix 2. So we
normalize the integral, multiplying by $-1/\pi$. This algorithm now
yields the charge of $-1$ for our electron model. Consider this a
calibration of the charge exhibited by the model. We will use this
same algorithm to assess the charge exhibited by all the other
particle models. One check we can apply at once is to see if the E
field, when projected onto the polar axis and integrated for a
complete cycle gives any result. The result is zero. So there is no
polar dipole associated with the E field. Henceforth we will use this
as our definition for model fields that are to be labeled E. We treat
the axial fields in a similar way, but obtain a different result. The
 field atop the ring , when projected onto the polar axis and
integrated around yields +$\pi$. The bottom field yields $-\pi$.
Both axial fields give zero for their integrated equatorial
projection. Thus they represent the field of a magnetic dipole, and
we label them B.

\subsection{The boson models} We now turn to the $Z^o$ model depicted
in the middle of Fig. 2 on the right. The gauge field components have
two nodes for once around the ring. That corresponds to one complete
period of the U(1) field, so we no longer have a spinor. 
Nevertheless, in Fig. 2 we show the boson fields going twice around;
this makes the resulting picture directly comparable to the fermion
depictions. What is special is that the twist direction reverses at
each node. Thus if we start with a counterclockwise twist (call it L)
in going from the node near the arrow, we change to a clockwise twist
(call it R) at the next node. We denote this twist pattern as LR. Had
we started noting the twist pattern at the next node, the pattern RL
would be obtained. The starting node is arbitrary so these two
patterns are equivalent. Both produce a neutral particle model. Our
algorithm for assessing the charge exhibited by these models has a
visual counterpart. If one takes a top view of the models, the fields
are projected onto the equatorial plane. One can then observe how the
+ and $-$ parts outside the ring add up. This is shown in Fig. 3. 

\FIGURE{
\epsfbox{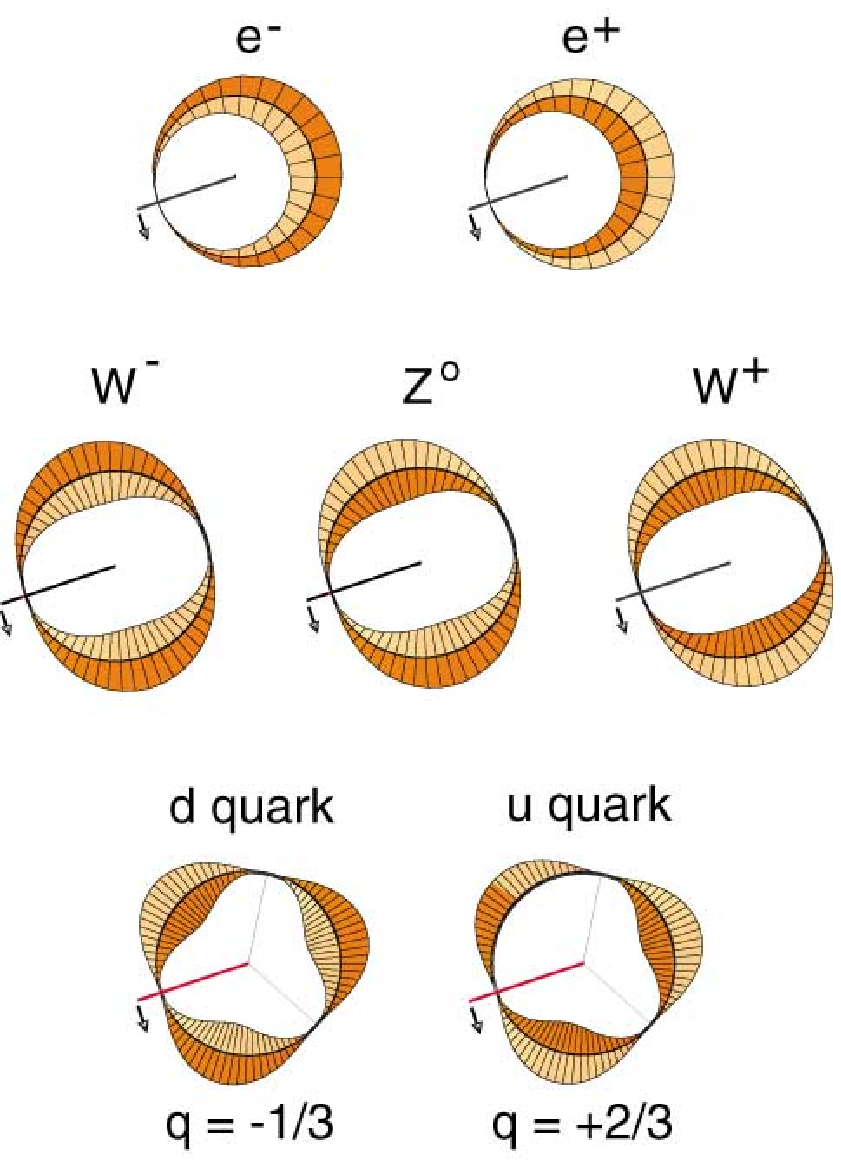}
\caption{Projection of particle
model E fields onto the equatorial plane. Distant observers see the
field integrated once around, but see only the portion external to
the dark circle - which represents an event horizon.}
}

For the $Z^o$ it can be seen that the $-$ field on one half of the
ring cancels the + field on the other half. In Fig. 3 the B fields
have been left out for clarity in being able to assess the electric
charges. Finishing out the massive bosons, we find that the LL twist
pattern makes the negative parts of the field add to give a model of
the $W^-$. And the RR twist pattern gives the $W^+$. The photon model
can conceptually be obtained from the $Z^o$ model by opening up the
latter at one of its nodes and pulling the ends out to plus and minus
infinity. This gives a model as shown on the middle left of Fig. 2.
Our photon model has a center of antisymmetry. This means that the
photon is its own antiparticle, a fact  consistent with the
experimental observation that a sufficiently energetic photon can
materialize into any particle-antiparticle pair. \subsection{The 
quark models} In the \textbf{n} = 3 particles, there are more ways
the twist degree of freedom can combine with the U(1) periodicity.
This produces more particle states, which shows up experimentally in
that quarks come in two charge states (the   + 2/3 \textbf{u} quark
and the $-$1/3 \textbf{d} quark), and in three colors. We are
challenged to model all these states in a consistent way within our
system. This can be done as follows. At the bottom right of Fig. 2 we
see the model of a \textbf{d} quark of some particular color. We know
it is a \textbf{d} quark from its electric charge of $-$1/3. That was
determined by our projection algorithm, visualized at the bottom of
Fig. 3. There for the \textbf{d} projection we see that the E field
is broken into three pieces by the nodes. Two of the pieces cancel,
leaving the third piece to provide the average field seen by distant
observers. How does this come about? As with the electron model, we
have to go around the ring twice to have the U(1) field advance an
integer number times 2$\pi$ so it can consistently repeat. So we have
a spinor wave function once again. The twist degree of freedom
therefore has six nodes at which it may reverse, or not. The twist
pattern that yielded the \textbf{d} quark of Figures 2 and 3 was LLR-LLR. Note
that there are two more equivalent patterns which also yield a $-$1/3
charge \textbf{d} quark. They are LRL-LRL and RLL-RLL. These patterns
are truly degenerate models in that they differ only in the origin
node one chooses for writing down the pattern. This is our model
representation of color. Call one pattern red, and the other two blue
and green. For an isolated quark, the phase denoted by color does not
matter. But when two or three quarks combine to form hadrons, their
relative phases do matter. It is then that color matters. The key
thing to notice about these \textbf{d} quark twist patterns is that
they are the same each time the excitation goes around the ring.
Suppose that on the second time around the ring, one of the twist
directions changed from what it was the first time around?  We get a
+2/3 charge \textbf{u} quark. The \textbf{u} quark projection shown
in Fig. 3 came from the twist pattern LLR-LLL. As can be seen, the E
field breaks up into three pieces at the nodes. But this time one
piece cancels itself out, leaving the other two to add up to 2/3 of a
charge. As before, we have three degenerate twist patterns that lead
to the same charge state. For the \textbf{u} quark, the other two are
LRL-LLL and RLL-LLL. Again we can assign these states color names.
These threefold possibilities for each charge state bring the
fundamental representation of the SU(3) group into our model.

\subsection{Gluons and Glucoms}
The n = 3 zero rest mass particle
shown at the lower corner of Fig. 2 does not seem to correspond to
any presently known particle. However, it may be considered to be a
gluon component in the following sense. In standard QCD, the gluons
are bicolored objects of spin one derived from the adjoint
representation \textbf{8} of SU(3). They can be considered to be made
up from two single colored objects from the defining representation
\textbf{3}  of SU(3). We call these gluon components glucoms. Our
unassigned n = 3 particle fits well as a representation of a gluon
component. The possible twist patterns LLR, LRL, and RLL represent
the three color states of the glucom. The corresponding anticolor
glucoms are RRL, RLR, and LRR. One gluon emission by a quark is then
equivalent to emission of a glucom and an antiglucom together. One
does not usually break gluons down this way in standard theory.
Nevertheless, this description seems to fit. In fact, we believe that
glucoms may one day be observed and found to play in important role
in some physical phenomena.

\section{Higher Generations of Fermion Families}
In model terms, how
shall we account for the experimental observation that the leptons
and the quarks come in three generations? The nature of the
excitation involved has been a puzzle of long standing \cite{Perl71}.
A hint comes from noting that in  the electron model the vector
fields are analogous to a closed twisting ribbon. Such ribbons have
another degree of freedom; they can writhe to make the centerline of
the ribbon non planar. So far, the electron model has a plane
circular propagation path with curvature, but no torsion. 
\FIGURE{
\epsfbox{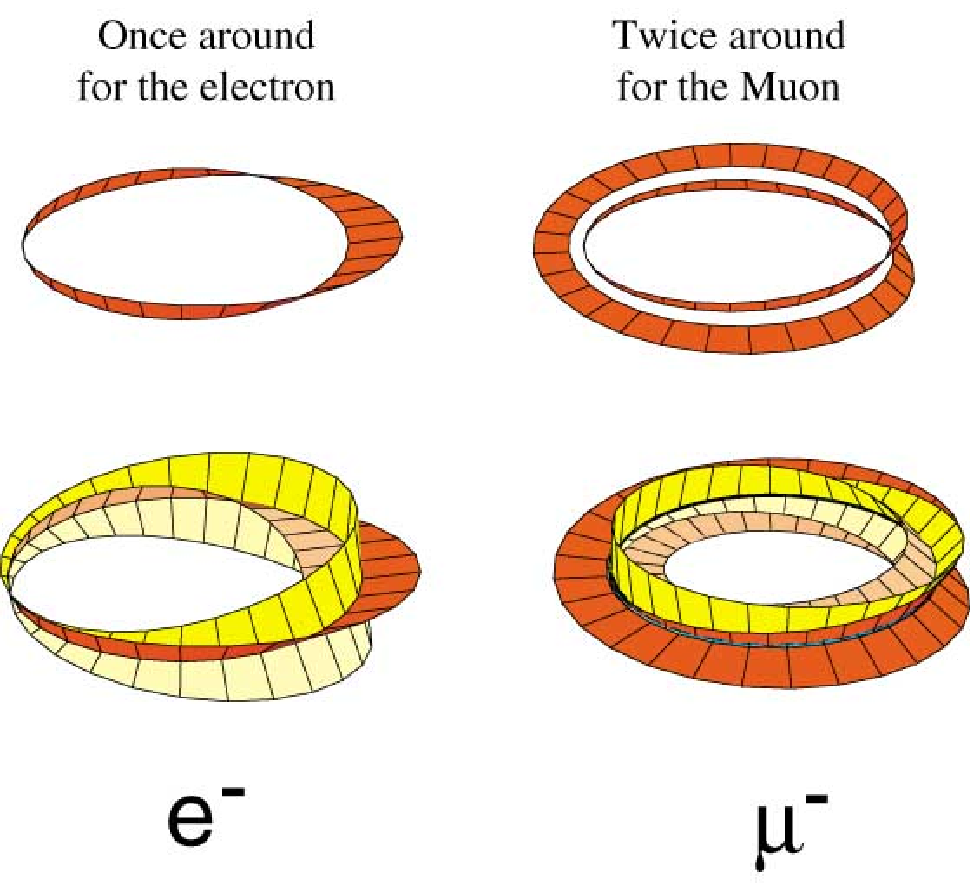}
\caption{The electron has a field
configuration with a half twist  for each $2\pi$ revolution of its
wave function. The muon has a  3/2 twist for each $2\pi$ revolution.
But by writhing back upon itself, this is equivalent to a  $4\pi$
revolution of the wave function with only a half twist. In this view,
 the muon  goes twice around,  giving a winding number of two. This
restores the  field geometries so they exhibit the same charge and
magnetic moment as  the electron. This makes the muon just a heavy
electron, as it is in nature.}
}
If at higher energy torsion became appreciable, then the propagation path
would begin to have non-zero writhe. We postulate that quantum
increases in writhe account for the excited states that higher
particle generations represent. Yet writhe seems not the proper
variable, as it is geometric, not topological. Closed ribbons in 3D obey a
simple relation: Lk = Tw + Wr \cite{Calu61}, where Lk is the linking
number of the two edges  considered as separate curves in space. For
a closed ribbon, Lk is a topological invariant. But the ribbon twist
(Tw) and writhe (Wr) can trade off with each other. However, for a
closed ribbon with a central obstruction, the writhe can no longer be
changed and it becomes topological. In such a case the writhe can be
given  a more physical representation and name; the writhe plus one
is the winding number, telling how many times the ribbon winds around
the obstruction before closing. This is a topological invariant which
seems a candidate for our generation quantum number. 
\FIGURE{
\epsfbox{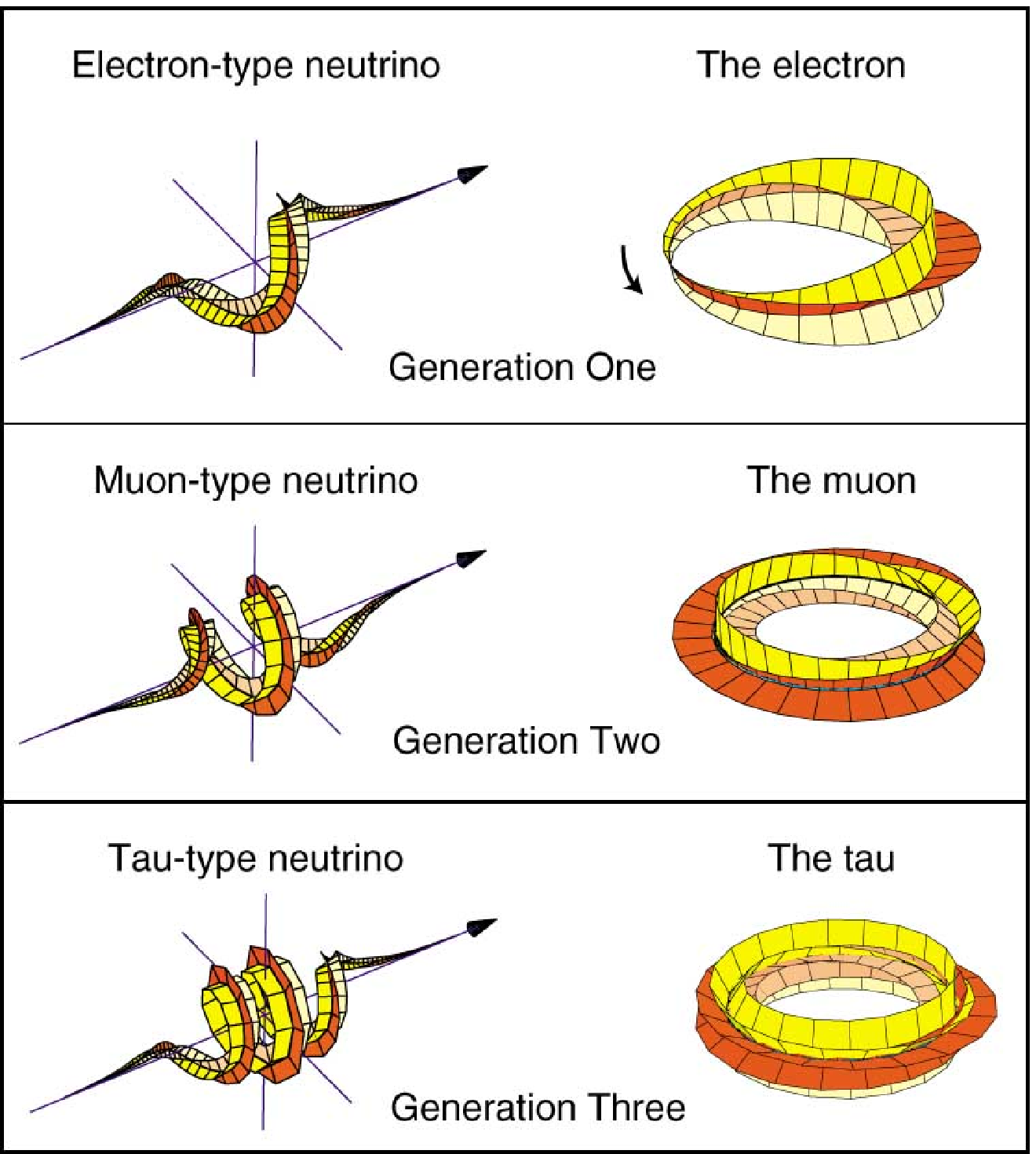}
\caption{Light front models for
three generations of  leptons. Generations one, two and three are
distinguished by  winding numbers of one, two and three respectively.
All the particles shown have one node and one half twist  in each
field component.}
}
We take the
interior of the horizon in our model to represent the interior of the
Kerr spacetime geometry - a special region, which is an obstruction around
which the field winds. Exploring this possibility, we assign a
winding number of one to the electron, two to the muon, and three to
the tau. The way this translates into model field geometries is shown
in Fig. 4 where we retain the one node and half twist  features of
the electron, but increase the winding number to two to obtain a muon
model. 

When we apply our algorithm of projecting and integrating the E field
around to compute the model charge, we find it comes out minus one,
just as for the electron. Increasing the winding number to three
yields a tau model that also exhibits a model charge of $-$1. This same
idea can be applied to the electron type neutrino model to yield the
corresponding second and third generation neutrinos. In Fig. 5 are
given depictions of the lepton particles for all three generations,
using the winding number approach. 

It is clear that the same idea may be applied to the quark models,
but we do not show the resulting models. In that case having three
nodes, the field geometries become so complex that the depictions are
confusing and not very useful. This approach to modeling the
generations of particles appears promising in that: (1) it is a
natural topological extension of our light front models, (2) it is
based upon a quantum number that is a topological invariant well
known in other contexts (quantum hall effects), (3) the models for
higher generation leptons exhibit the same electrical charge as the
first generation models, and (4) it uses a mechanism that works
equally well for the leptons and the quarks. But there is a more
compelling reason to believe that there is some truth in this
description. Witten [13] has shown that the expectation values of
observables depend upon the writhe of framed links in a topological
field theory derived from a Chern-Simons action. When we plot the
lepton masses for all three generations in that way (see below), we
obtain a beautiful linear plot whose parameters are given by the weak
coupling constant $e^2$, and the weak mass scale \textbf{$m_W$}. No
adjustable parameters are needed. This is discussed below.

\section{Quantum Numbers}
We have introduced several quantum numbers
that in model terms distinguish one elementary particle from another.
To bring them into focus, we summarize them and their significance in
Table 1.

\TABLE{
Table 1. Quantum numbers for specifying the different elementary
particles.
\begin{tabular}
[b]{|c|c|c|p{6.8cm}|}
\multicolumn{4}{c}{\textbf{Quantum Numbers for All Particles}}  \\
\hline \textbf{Symbol}&\textbf{Allowed \newline
Values}&\textbf{Name}&\textbf{Significance} \\ \hline \textbf{b} &
0,1 & boundary type & b = 0 for zero rest mass particles \newline b =
1 for particles with rest mass \\ \hline \textbf{h} & +1, 0, $-$1 &
handedness &  h = +1 for particles \newline h =  \hspace{0.2cm}0 for
self-conjugate particles \newline h = $-$1 for antiparticles \\ \hline
\textbf{n} & 1,2,3 & nodes & n = 1 for leptons \newline n = 2 for
gauge bosons \newline n = 3 for quarks and glucoms \\ \hline
$\mathbf{n_w}$ & 1,2,3 & winding number & $\mathrm{n_w}$ = 1 for 1st generation
particles \newline $\mathrm{n_w}$ = 2 for 2nd generation particles \newline $\mathrm{n_w}$ = 3
for 3rd generation particles \\ \hline
\multicolumn{4}{c}{\textbf{Quantum Numbers for $n>1$ Particles}} \\
\hline \textbf{t} & +1, $-$1 & twist splitting & t = +1 and n = 2
gives W particles \newline t = \hspace{0.1cm}$-$1 and n = 2 gives the
$Z^o$ \newline t = +1 and n = 3 gives u, c, or t quarks \newline t =
\hspace{0.1cm}$-$1 and n = 3 gives d, s, or b quarks \\ \hline
\multicolumn{4}{c}{\textbf{Quantum Numbers for $n>2$ Particles}} \\
\hline \textbf{c} & 1,2,3 & color& h = \hspace{0.2cm}1 and c = 1
gives red \newline h = \hspace{0.2cm}1 and c = 2 gives green \newline
h = \hspace{0.2cm}1 and c = 3 gives blue \newline h = $-$1 and c = 1
gives antired \newline h = $-$1 and c = 2 gives antigreen \newline h =
$-$1 and c = 3 gives antiblue \newline \\ \hline
\end{tabular}
}

A dream of ours has been to have a grand computer program that would
take values of these quantum numbers as input, and then produce as
output a light front model depiction of the particle specified. In
fact, we find it more convenient to have several programs (one for
the massless particles, another for the massive particles, etc.) 
Still, the dream challenges us to come up with a minimal set of
quantum numbers that can uniquely specify each of the known
particles. In addition we seek quantum numbers that have an obvious
physical, geometric or topological meaning. Here is our set.  First
we choose a quantum number \textbf{b} whose function is to specify
the type of boundary condition the wave function must satisfy. Its
value determines whether our chosen particle does or does not have
rest mass. Next we specify a chirality or handedness factor
\textbf{h}; it specifies whether we are dealing with a particle or
antiparticle. The special value \textbf{h} = 0 is used for particles
like the photon that are their own antiparticle. The choice between
leptons, bosons, and quarks is made by specifying \textbf{n}, the
number of nodes in the gauge wave function. Lastly we choose the
particle generation by giving the winding number $\mathbf{n_w}$  (note
that bosons have only the $\mathbf{n_w}$ = 1 generation for reasons
discussed below). These are the basic quantum numbers that apply to
all the particles. When \textbf{n} $>$ 1, we need a quantum number to
account for the splitting between the W and $Z^o$, and also between
the \textbf{u} and \textbf{d} quark sectors. In the models, the
feature which distinguishes these particles is the total twist of the
gauge field for one complete cycle of the U(1) field. So we define a
twist splitting quantum number \textbf{t}. In a similar manner, when
\textbf{n} $>$ 2, we need a quantum number to specify the twist
permutation pattern. So for the quarks we define a color quantum
number \textbf{c}. The allowed values for these quantum numbers are
given in Table 1. We not only use these quantum numbers for
specifying particles to our programs, but also will use them in the
particle mass relationships discussed below.

\section{Particle Mass Relationships}
The masses of the three
generations of leptons are well known from experiment ($m_e$ = 0.511
MeV, $m_\mu$ = 105.7 MeV, and $m_\tau$ = 1777 MeV) \cite{Part96}, but
they have never been understood theoretically. It is clear that the
masses rise rapidly and perhaps exponentially with generation number.
We, and many others, have tried fitting them with an exponential
relationship. See reference \cite{Ruch94} for a recent attempt. The
results of those attempts have never really been satisfactory. The
trouble was that no one had any idea what kind of excitation was
involved in going from the first generation to the second or third.
But in the context of the light front models, we do have an idea. It
is the increasing writhe (or winding number) that leads to the higher
generations. With that as a hint, we ask how such a thing might arise
in quantum field theory.

Our models begin in 3 + 1 dimensional spacetime. But by assumption I,
the gauge excitations are restricted to be transverse. This reduces
their spacetimes to 2 + 1 dimensions. So the theory we seek should be
a 2 + 1 dimensional quantum field theory whose observables (like rest
mass) are functions of topological invariants. The natural choice is
a Chern-Simons (CS) topological quantum field theory (TQFT), as
wonderfully explicated by Witten \cite{Witt89}. We are further
encouraged in this direction when the writhe and winding number of
the gauge fields emerge from such a theory as a key variables.

To apply CS theory, one must choose an oriented three manifold and a
gauge group. For application to our models, we choose $R^3$ for the
manifold and U(1) as the gauge group. (Recall that the expected SU(2)
group for fermions enters the models later.) These choices lead to
the simplest possible CS action:

\begin{displaymath} \mbox{$S_{CS}$}
= \frac{k}{4\pi g^2} \int _{R^3} A \wedge \,dA
\end{displaymath}
where A is the vector potential, k is the level (which we will set to
one to obtain the lowest state), and g is the coupling constant. For
the case of a transverse U(1) vector field propagating around an
$S^1$ path embedded in $R^3$ with blackboard framing (no twists), the
CS action  equates to the Gauss' self-linking number or writhe of the
field multiplied by the scale factor. This is shown explicitly in
\cite{Baez94}. Witten \cite{Witt89} went farther, showing that the 
Feynman path integral of the Chern-Simons action over all gauge
orbits around the $S^1$ link C yields the partition function of the
theory - a topological invariant. Next one defines a Wilson line
operator $W$ which gives the holonomy of the connection integrated
around the cyclic path:

\begin{displaymath} \mbox{W} = \mbox{P} exp[
\int _{C} A \: dA] \end{displaymath} Combining these in the path
integral, Witten showed \begin{displaymath} \langle \mbox{vev}
\rangle \: = \int DA\:  exp \:[\mathrm{i}S_{CS}] \:\prod_i\: W_i =
e^{\Lambda \cdot \mathrm{writhe}} \:[\mbox{Jones Polynomial}(C_i)]
\end{displaymath}

The result is a topological invariant - the vacuum
expectation value (vev) of the operator $W = \prod_i\: W_i$. (We have
omitted from the equation above the partition function that serves as
a normalizing factor. As will be seen below, we let the data plot
determine it, which in our case becomes the mass scale factor.)

Witten showed that this vev is also given by the well known Jones
polynomial invariant $\langle \mbox{J} \rangle$ for the link
multiplied by a writhe factor that takes into account the
conformation of the embedding of the framed $S^1$ link in $R^3$. In
knot theory, this combination goes under the name of the Kauffman
bracket for certain  choices of its variable.

We hope to apply this to the present models by considering the
experimental rest masses of the $e$, $\mu$, and $\tau$ leptons as
observables depending upon  some function of the vev  of the
applicable Chern-Simons TQFT. But there are subtle features to be
dealt with.

First, the framed link invariant given by the CS path integral is
$\langle \mbox{K} \rangle$, the Kauffman Bracket \cite{Kauf91}, whose
framing dependence is made explicit as a writhe prefactor times the
Jones polynomial, thus $\langle \mbox{K} \rangle \: =
(q)^{\mathrm{writhe}} \langle \mbox{J} \rangle$. The quantity q is
the Chern-Simons phase $exp(i\pi/k)$ for the U(1) case. For the level
k = 1, q = $-1$. Next, there is an issue of signs \cite{Sawi96}.  To
give the vev,  the Kauffman bracket must be multiplied by
$(-1)^{\mathrm{winding\:number}}$. As was mentioned earlier, for
planar projections, the winding number is just the $(\mathrm{writhe}
+ 1)$. Thus we get the curious equation:

\begin{eqnarray*} \langle
\mbox{vev} \rangle \: & = & (-1)^{\mathrm{writhe}}  \cdot
(-1)^{\mathrm{writhe} + 1}\prod_i \: \langle J_i \rangle \nonumber\\
& = & - \prod_i\: \langle J_i \rangle
\end{eqnarray*}

The surprising effect of all this is that the writhe
prefactor arising from the CS action has no effect for our level k=1
U(1) gauge field other than to introduce a minus sign. What is left
in the vev is the product of Jones Polynomials, one  for each loop
the field makes. This is relevant since for the electron model we go
once  around; for the mu model, twice around, and three times for the
tau. These Jones Polynomials are usually normalized to unity for the
unknot. But they arise from the Wilson loop operators. In our fermion
case, due to the half twist of the fields as they go once around,
they have  traversed only half of the field cycle. If we normalize to
a complete field cycle, then each loop yields a holonomy of 1/2. In
our model for the generations, the winding number of each model gives
the number of loops;  it also gives the number of factors of 1/2 in
the vev. By this argument, we obtain

\begin{eqnarray} \langle
\mbox{vev} \rangle \: & = & -\left(\frac {1}{2}\right)^{n_w}
\end{eqnarray}
where $n_w$ is the winding number.

A last complication is that this TQFT Expectation Value has been
derived for the 2 + 1 spacetime of our models on the light cone.
 But we make our observations asymptotically far away in a
flat 3 + 1 spacetime. There must be a mapping from one to the other
that makes our observable (we focus on mass) some function of the
TQFT Expectation Value. We take our clue from Hawking and Ellis \cite{Hawk76},
who prove that timelike paths from asymptotic infinity to the light cone
constitute an exponential map. If we
assume an exponential relationship, then we can write a possible
lepton mass relation as
\FIGURE{
\epsfbox{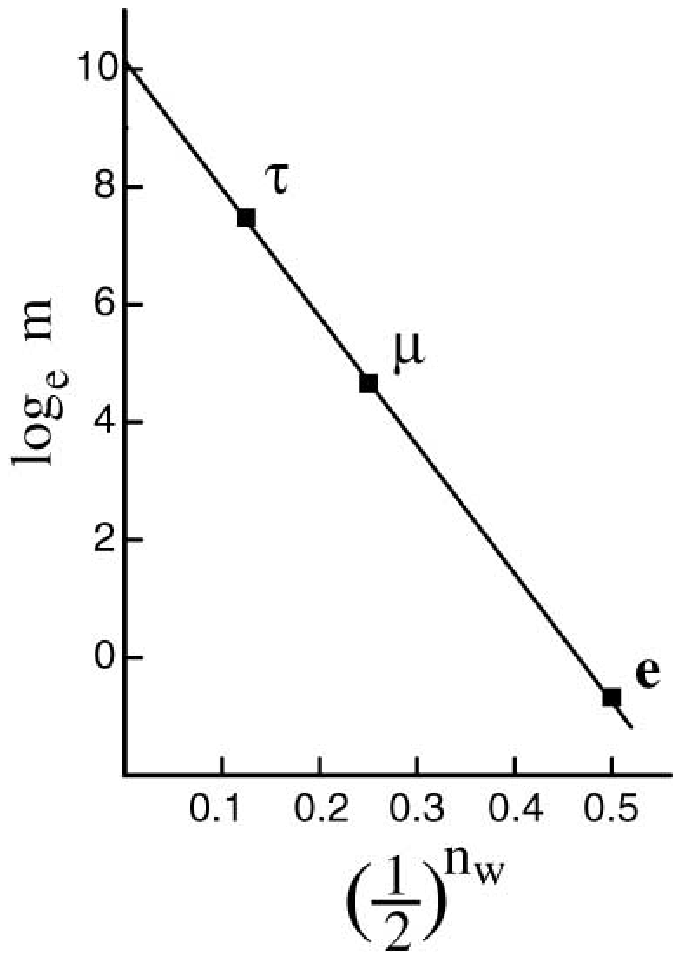}
\caption{Mass plot for the leptons.
On the  abcissa $n_w$ is the gauge field winding number with $\mathbf{n_w}$ =
1, 2, 3 for the  first, second, and third generation particles
respectively. All  masses are in MeV. The line is theoretical from
equation (4). It has a  slope of $-$2/(4$\pi \alpha$) and an intercept
of log ($m_w$/$\pi$). where $m_w$  is the mass of the W boson in
MeV.}
}
\begin{equation} \frac{m}{m_0}  = 
exp\left[-a\left(\frac{1}{2}\right)^{n_w}\right]
\end{equation}
or, in logarithmic form
\begin{equation} \log m  = 
-a\left(\frac{1}{2}\right)^{n_w} + \log m_0
\end{equation}
where $m_0$ is the mass scale factor, $a$ incorporates the coupling
constant of the CS action, and $n_w$ is the winding number. The
validity of this functional form can be tested by plotting $\log m$
versus $\left(\frac{1}{2}\right)^{n_w}$ with $n_w$ = 1, 2, and 3 for
the $e$, $\mu$, and $\tau$ leptons respectively. Those winding number
values are derived from our light front models of the lepton
generations as was described in Section 5. If the functional form is
correct, we should get a linear plot whose slope is $-a$ and whose
intercept gives $\log m_0$. We have done this (expressing all masses
in MeV) as shown in Fig. 6. 

The result is a good linear plot. Surprisingly, the value of $a$ from
the slope is equal to 2/(4$\pi\alpha$), where $\alpha$ is the fine
structure constant (1/137). In natural units, 1/(4$\pi\alpha$) is
$e^2$, the electromagnetic coupling constant. In like fashion, the
plot intercept gives $m_0 = \frac{m_W}{\pi}$ where $m_W$ is the mass
of the W boson (80,430 MeV). Using these values for $a$ and $m_0$,
gives the lepton mass formula

\begin{equation} m_i  = 
\left(\frac{m_W}{\pi}\right)exp\left[-\frac{2}
{4\pi\alpha}\left(\frac{1}{2}\right)^{n_{w_i}}\right]
\end{equation}
where for $i=1,2,3$ the winding number $n_{w_i} =1,2,3$ for the $e$,
$\mu$, and $\tau$ leptons respectively.

The line plotted in Fig. 6 is that given by equation (7.4). Its
goodness of fit can be judged by the calculated lepton masses: $m_e$
= 0.47 MeV (7.8\% low), $m_\mu$ = 110 MeV (3.8\% high), $m_\tau$ =
1677 MeV (5.6\% low). If we count $\alpha$ and $m_W$ as relevant
constants of nature, then this representation has been obtained with
no adjustable parameters. We take the linear fit, and the appearance
of the natural constants as indicators that our approach is on the
right track. This kind of relationship can be extended to include 
the quarks, but those cases are more complex and require  new
approximations. That work will be described in a separate paper.\cite{Mill}

We should comment upon the gauge bosons, however. Experimentally, the
bosons do not show generational behavior. Why not? In the context of
our CS TQFT picture let us focus on the Wilson loops for the bosons.
They  integrate over one complete cycle of the field in going once
around the loop. The positive half cycle cancels the negative half
cycle giving a zero result. Thus we replace the 1/2 in equation (1)
with zero. The result for bosons is that their vevs no longer depend
upon the winding number.

\section {Spacetime Mappings} Several times above, we have alluded to
the need for mapping expectation  values that arise in the vicinity
of particle horizons onto the spacetime of distant real observers.
For measuring electric charge, distant observer measurements must
time-average over the cyclic motion of the fields. But  there is a
spatial mapping that is not trivial. If electrons are really tiny,
rapidly rotating black holes of Kerr geometry as our models
postulate, then the  spacetime just outside their horizon is dragged
into rotation as well. Also, the E field leaves the electron
predominently in its equatorial plane. How can it be that a distant,
massive observer sees a  nonrotating radial field from what appears
as a point charge? We depict this problem in Fig. 7, using a cartoon
drawing.

\FIGURE{
\epsfbox{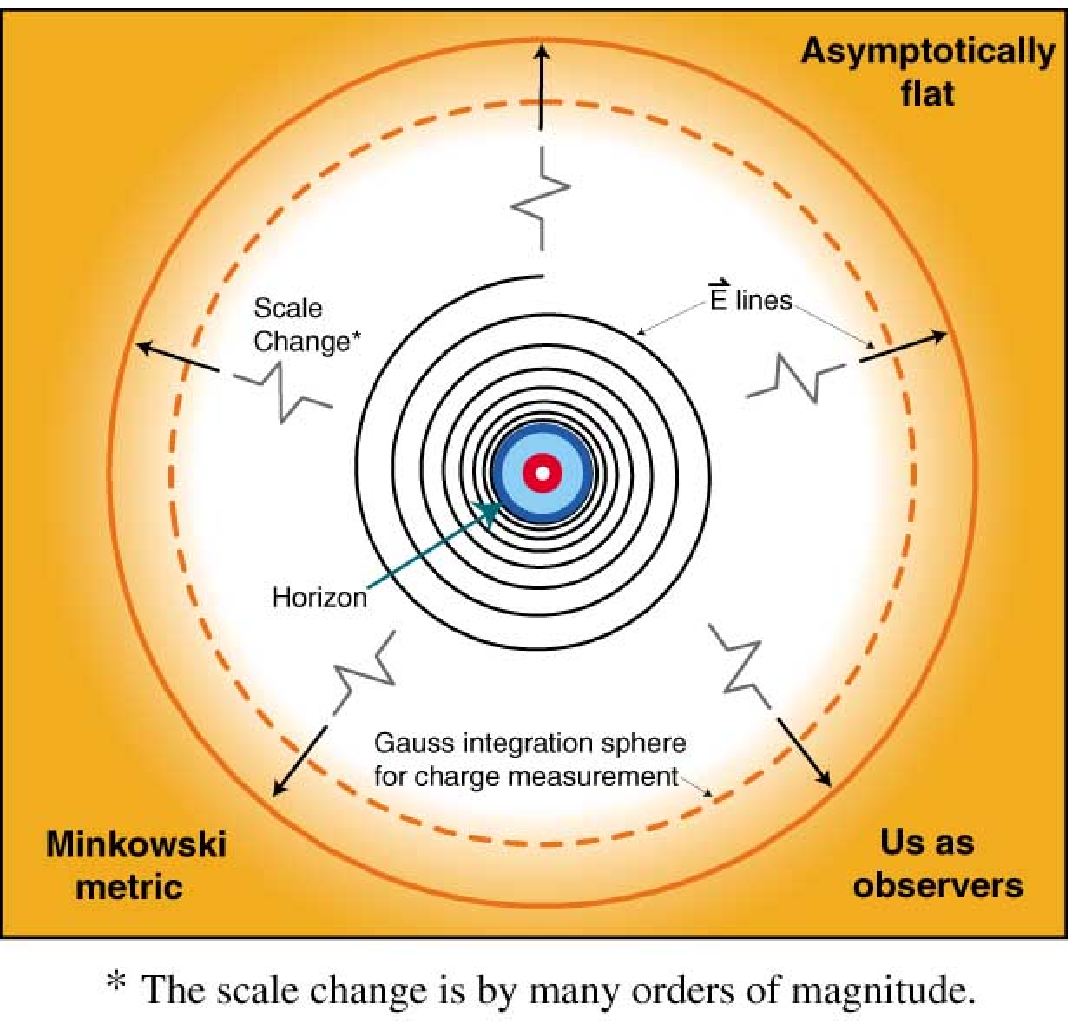}
\caption{A cartoon depiction of a
static mapping  from the Kerr geometry of the electron E field onto
the Minkowski geometry  of an observer asymptotically far away. We
are looking down onto the equatorial  plane of the particle.}
}

The way rotating fields can map to stationary points  has an answer
long known in physics. It is the spinor property of entangling its
fields upon a 2$\pi$ rotation and then untangling them upon a further
2$\pi$ rotation. This is nicely illustrated by Misner, Thorne and
Wheeler in \cite{Misn73}. There is even a U. S. Patent \cite{Adam71}
for a device that accomplishes this trick on a continuous basis. But
it works only for spinors such as fermions. Bosons like the W and the
$Z^0$ are not spinors. What about them? This leads us to a
speculation about a troubling observation.

Why are the first generation fermions very light while the simplest
massive bosons very heavy? The electron is extremely light, with mass
of about .5 MeV. The u and d quarks have mass less than 10 MeV. In
contrast, the W boson has a mass of over 80,000 MeV. The difference
cannot be ascribed to the difference in node numbers, for the light
electron and quarks have node numbers of one and three respectively,
while the heavy bosons have node number two.  We speculate that it is
the mapping of fields propagating to distant observers that gives
rise to the huge difference in observed masses. All of the particles,
as modelled here, are sources of twisting gauge fields propagating
outward. These fields carry energy and contribute to the energy
momentum tensor that distant observers use to measure mass. For the
fermions, the twist energy gets averaged over only a 4$\pi$ rotation
of the model fields due to the 4$\pi$ symmetry property of spinor
fields. But there is no such limited averaging for the bosons. They
do not possess the spinor symmetry property. For the bosons, the
twist gauge field energy gets averaged over a large number of model
rotation cycles. How to specify this is not clear. Yet the
qualitative difference between the fermion and boson cases is clear.
We will not pursue this farther here.

\section {Discussion}

Now that the models have been developed and presented, it seems a
good time to reconsider some of the assumptions that went into them.
As illustrated in Fig. 1, we postulated that the propagating gauge
fields would writhe about the geodesic propagation path in regions
where the excitation was intense. How might this arise? One can
always speculate about nonlinearities in the field equations but we
can be more specific. We invoked a Chern-Simons term in the action
for a TQFT to explain the mass plot of Fig. 6. Physically considered,
the CS form is a generator of torsion in the following sense. A
propagating gauge field sweeps out a space curve whose  direction at
a point is given by the tangent to the curve. The gradient dA defines
the normal to the curve at the same point. This establishes a plane
in which the space curve may exhibit curvature. The wedge product
with A defines the binormal direction which is perpendicular to the
plane just described. Motion of the propagating field in the binormal
direction exhibits torsion. In view of this, we ascribe the torsion
of the paths shown in Fig. 1 to a CS term in the Lagrangian. In fact,
the CS term must be the dominant term in the Lagrangian in view of
the excellent fit to the lepton masses shown in Fig. 6. The fit is
not perfect, however. This suggests to us that there are Maxwell
terms and perhaps others we have ignored. There may also be quantum
corrections we have left out. Nevertheless, it appears that the
Chern-Simons term dominates.

As the light front models were being developed, it became clear that
the twisting of the fields about their propagation direction was a
key feature. First for the electron model, and then for the bosons
and quarks the change of twist by $\pi$ radians from one node to  the
next was required to give reasonable model geometries. Since this
appeared in every case, we have elevated it to the status of a
natural law as stated in assumption IV. Perhaps the twist degree of
freedom should be made the basis of the gauge field. It exhibits U(1)
group behavior. And the electromagnetic field could be considered to
arise as the  gradient of the $\theta$ field. The fact that CS TQFT
assigns an important role to the twist of framed links representing
the gauge fields also suggests one should pursue these possibilities.

The modeling of leptons, bosons and quarks as 1, 2 and 3 node 
harmonic oscillator states of the gauge field immediately suggests
that states with 4, 5, and 6 or more nodes might be made with
accelerators of higher energy than those presently available. This
theory, unlike some, does not predict a desert unpopulated with
states until one approaches the Planck energy. The n = 3 states
(quarks) show a much richer phenomenology than the n = 1 states
(leptons). One might expect the n = 5 states to be richer yet.

These models cast new light upon whether magnetic monopoles can exist
or not. In looking at the electron model of Fig. 2 it would seem that
a particle with magnetic charge could be made by merely rotating all
the fields $\pi$/2 radians about their propagation direction. Then
the  yellow (light grey) fields would become radial and the orange
(dark grey) fields would become polar. Yet from a distant observers
point of view, nothing has changed. He still sees a particle with a
radial gauge flux that looks like an electric field. And he sees a
dipolar field that looks magnetic. The point is that the orthogonal
components of the gauge field are completely equivalent in the
massless particles. In the massive particles they become different
because of the symmetries they are forced to adopt. We label the
radial field E and the polar field B. Given this viewpoint, our
models suggest that magnetic monopoles cannot be made. This is in
keeping with the experimental fact that they have never been observed
in a repeatable way despite many searches. 

As was shown in Fig. 2, we had to postulate an as yet unknown
particle that is the n = 3 counterpart to the neutrino. We 
tentatively called it a glucom as it appears as a gluon component. We
suggest that this is a real particle yet to be discovered. One can
speculatively think of several places where glucoms might play a role
in physics and cosmology. Suppose that in the core of the sun some of
the newly formed neutrinos could combine with photons to make
glucoms. They likely would not be detected by present solar neutrino
detectors. This could contribute to the observed solar neutrino
deficit. Glucoms could also make up some of the dark matter of the
universe. We will not pursue these speculations here.

The elementary particles as modeled here represent a new and lower
level in our understanding of the structure of matter. It appears to
be a level with many interesting structures that are both geometric
and topological in nature. We suggest that the structures be called
geotopes.  One can only hope that geotopic structure theory will need
to be extended to handle the n = 4, 5 and 6 geotopes.

\acknowledgments
 The author thanks W. E. Palke for many helpful
discussions. 

\section{Appendix 1. \emph{Mathematica} Program for the\\
Quark and Electron Light Front
Models}

\begin{verbatim}

(* This program depicts a light front model of the d quark with the
setting nodes = 3. It can also depict the electron model if the value
is changed to nodes = 1.

The model is specified using toroidal coordinates of scale
baseRadius.  There are two plotting parameters, t and v. The angular
parameter is t, with range {t,0,4Pi}. We make the angular parameter
go twice around to accommodate spinor wave functions. The radial
parameter is v, with range {v, 0, maxMinorRadius}. The positive and
negative parts of the E field are rendered using light orange and
intense orange colors respectively. The N and S parts of the B field
use yellow and light yellow. In toroidal coordinates, the fields are
defined by:

Plus E field  (orange):           North B field ( yellow): 
r = v Sin[nodes*t/2]              r = v Sin[nodes*t/2] 
theta =twists*t                   theta = twists*t/2 + Pi/2
phi = t                           phi = t

Set pcFactor to +1 for a particle, or to -1 for an antiparticle *)

nodes = 3; twists = nodes/2; pcFactor = +1;

baseRadius = 3.0; maxMinorRadius = 1.0;

If[nodes==1, ch=+1, ch=-1]; (* Sets correct chirality for electron
case *)

ToroidalToXYZ[chirality_,r_,theta_,phi_] [hueColor_,saturation_] :=
  {(baseRadius + r Sin[-chirality*pcFactor*theta]) Cos[phi],
  (baseRadius + r Sin[-chirality*pcFactor*theta]) Sin[phi], r
  Cos[-chirality*pcFactor*theta], SurfaceColor[Hue[hueColor,
  saturation, 1.0]]}

Field[chirality_,sign_,start_,end_,orient_,hue_,sat_] :=
  ParametricPlot3D[ ToroidalToXYZ
  [chirality,v*(sign)*Sin[nodes*t/2],twists*(t+orient),t] [hue,
  sat]//Evaluate, {t,start, end},{v,0.019,maxMinorRadius}, Boxed ->
  False, PlotPoints->{30,2}, ViewPoint->{1.087, 4.108, 2.2},
  LightSources -> {{{1.,0.,1.}, RGBColor[1,1,1]}, {{1.,1.,0},
  RGBColor[1,1,1]}, {{0.,1.,1.}, RGBColor[1,1,1]}, {{-1.,-1.,1.},
  RGBColor[1,1,1]}}, DisplayFunction->Identity];

orange1 =     Field[+1, -1, 0,     2Pi/3, 0, .03, 1]; 
orangelite1 = Field[+1, +1, 0,     2Pi/3, 0, .03, 0.75]; 
orange2 =     Field[+1, -1, 2Pi/3, 4Pi/3, 0, .03, 1]; 
orangelite2 = Field[+1, +1, 2Pi/3, 4Pi/3, 0, .03, 0.75]; 
orange3 =     Field[ch, -1, 4Pi/3, 2Pi,   0, .03, 1]; 
orangelite3 = Field[ch, +1, 4Pi/3, 2Pi,   0, .03, 0.75];

OrangeField = Show [orangelite1,orangelite2,orangelite3,
  orange1,orange2,orange3, DisplayFunction->Identity];

yellow1 =     Field[+1, -1, 0,     2Pi/3, Pi, 0.2, 1]; 
yellowlite1 = Field[+1, +1, 0,     2Pi/3, Pi, 0.2, 0.75]; 
yellow2 =     Field[+1, -1, 2Pi/3, 4Pi/3, Pi, 0.2, 1]; 
yellowlite2 = Field[+1, +1, 2Pi/3, 4Pi/3, Pi, 0.2, 0.75]; 
yellow3 =     Field[ch, -1, 4Pi/3, 2Pi,   Pi, 0.2, 1]; 
yellowlite3 = Field[ch, +1, 4Pi/3, 2Pi,   Pi, 0.2, 0.75];

YellowField = Show[yellowlite1,yellowlite2,yellowlite3,
  yellow1,yellow2,yellow3, DisplayFunction->Identity];

axis = Graphics3D[Line[{{0,0,-5},{0,0,4}}]];

redLine = Graphics3D[ {RGBColor[1,0,0], Thickness[.01],
  Line[{{0,0,0},{3.8,0,0}}]}];

Particle = Show[OrangeField,YellowField,axis,redLine,
  ViewPoint->{1.087, 4.108, 2.2}, Axes->False, PlotRange->All,
  DisplayFunction->$DisplayFunction]

(* Graphic deleted. See Fig. 2 at the bottom right for depiction. *)

(* Top view for charge evaluation *) 
Show[OrangeField,redLine,
  ViewPoint->{-0.000, 0.059, -3.383}, Axes->False,
  DisplayFunction->$DisplayFunction]

(* Graphic deleted. See Fig. 3 at the bottom left for depiction. *)

\end{verbatim}

\section{Appendix 2. Electric Charge Calculation\\
for the Particle Models}

In this appendix are given explicit calculations for the particles
shown in Fig. 3 whose charges were deduced graphically as described
in the text.

The algorithm for deriving the electric charge that a given particle
model exhibits to distant real observers is as follows: For twice
around the ring, project the outwardly  directed E field vectors onto
the equatorial plane of the ring and integrate. Normalize by
multiplying by $-\pi^{-1}$. As shown below, this fixes the charge on
the electron model at $-1$.

\textbf{Electron Model Charge}

From Appendix 1, the length of the orange E field vector is
$\sin(t/2)$. The orangelite (or positive) part of the field points
inward and is hidden behind the horizon. So it is not included in the
integral. The angle theta is measured down from the axis of the ring,
and is the complement of the projection angle we want. So we project
using $\cos(\pi/2 -\mbox{nodes}*t/2) = \sin(\mbox{nodes}*t/2)$. Hence
the charge calculation is:

\begin{displaymath} \mbox{Charge} = -\frac{1}{\pi} \int ^{2\pi} _0
\sin(t/2) \sin(t/2) \,dt = -1 \end{displaymath}

\textbf{Z$^0$ Model Charge}

From Fig. 3, we see that for the first time around the ring, we
integrate negative E field vectors over the range 0 to $\pi$, and the
positive E field vectors from $\pi$ to $2\pi$. The second  time
around all the vectors are internal and make no contribution.

\begin{displaymath} \mbox{Charge} = -\frac{1}{\pi} \int ^{\pi} _0
\sin^2(t) \,dt + \frac{1}{\pi} \int ^{2\pi} _{\pi} \sin^2(t) \,dt = 0
\end{displaymath}

\textbf{W$^-$ Model Charge}

From Fig. 3, we see that there are two negative E field vector
contributions external to the ring:

\begin{displaymath} \mbox{Charge} = -\frac{1}{\pi} \int ^{\pi} _0
\sin^2(t) \,dt - \frac{1}{\pi} \int ^{4\pi} _{3\pi} \sin^2(t) \,dt =
-1 \end{displaymath}

\textbf{d Quark Model Charge}

With three nodes and going twice around, we have six pieces of E
field to consider. From the projection of Fig. 3, it is evident that
only three of them are external to the ring. Identifying those using
the program of Appendix 1, we have:

\begin{eqnarray*} \mbox{Charge} &=&-\frac{1}{\pi} \int ^{2\pi/3} _0
\sin^2(3t/2) \,dt \\ & & -\frac{1}{\pi} \int ^{10\pi/3} _{8\pi/3}
\sin^2(3t/2) \,dt \\ & & +\frac{1}{\pi} \int ^{4\pi} _{10\pi/3}
\sin^2(3t/2) \,dt = -\frac{1}{3} \end{eqnarray*}

\textbf{u Quark Model Charge}

The u quark differs from the d quark in having four pieces of the E
field external as can be seen from Fig. 3. So there is a fourth term
in the charge calculation:

\begin{eqnarray*} \mbox{Charge} &=&\frac{1}{\pi} \int ^{2\pi/3} _0
\sin^2(3t/2) \,dt \\   & & +\frac{1}{\pi} \int ^{2\pi} _{4\pi/3}
\sin^2(3t/2) \,dt \\ & & +\frac{1}{\pi} \int ^{10\pi/3} _{8\pi/3}
\sin^2(3t/2) \,dt \\ & & -\frac{1}{\pi} \int ^{4\pi} _{10\pi/3}
\sin^2(3t/2) \,dt = +\frac{2}{3} \end{eqnarray*}


\begin{thebibliography}{18}
\bibitem{Skyr61} T. H. R. Skyrme, \emph{A non-linear theory of strong interactions}, {\it Proc.
Roy. Soc.} {\bf A262} (1958) 260 ; \emph{Particle states of a quantized meson field}, ibid. 
{\bf A262} (1961)  237; \emph{A Unified Field Theory of Mesons and Baryons},
{\it Nucl. Phys.} {\bf 31} (1962) 556.

\bibitem{Thir58} See the discussion of the Thirring model by
S. Coleman,  {\em Aspects of Symmetry\/} (Cambridge University Press,
Cambridge, 1985) p. 246ff.

\bibitem{Enz77} U. Enz, \emph{A new type of soliton with particle properties}, 
{\it J. Math. Phys.} {\bf 18} (1977) 347 ; \emph{A particle model based on stringlike solitons},
ibid. {\bf 19} (1978) 1304.

\bibitem{Sen95} A. Sen,  \emph{Extremal black holes and elementary string states},
{\it Mod. Phys. Lett.} {\bf A10} (1995) 2081.

\bibitem{Klei97} B. Kleihaus and J. Kunz, \emph{Static axially summetric
Einstein-Yang-Mills dilaton solutions: Regular solutions}, {\it Phys. Rev.} {\bf D57} (1998) 834;
\emph{Static axially summetric Einstein-Yang-Mills dilaton solutions. II. Black-hole
solutions}, ibid. {\bf D57} (1998) 6138.

\bibitem{Witt88} E. Witten, \emph{Topological quantum field theory}, {\it Commun.
Math. Phys.} {\bf 117} (1988) 353.

\bibitem{Gamb96} R. Gambini and J. Pullin, {\em Loops, Knots, Gauge
Theories and Quantum Gravity\/} (Cambridge University Press,
Cambridge, 1996).

\bibitem{Cast97} A. De Castro and A. Restuccia, \emph{Topologically Massive
Models from Higgs Mechanism}, hep-th/9706060.

\bibitem{Perl71} M. L. Perl, \emph{How does the muon differ from the
electron?}, {\it Physics Today} (July 1971) p. 34; \emph{The Leptons after 100
Years}, ibid. (October 1997) p. 34.

\bibitem{Calu61} G. Calugareanu, \emph{Sur les classes d'isotopie
des noeuds tridimensionnels et leur invariants}, {\it Czechoslovak Math J.} {\bf 11} (1961)
588-625; J. White, \emph{Self-linking and the Gauss integral in higher dimensions},
{\it Amer. J. Math.} {\bf 91} (1969) 693-728; F. B. Fuller,
\emph{The writhing number of a space curve}, {\it Nat. Acad. Sci. USA}  {\bf 68} (1971) 815-819.

\bibitem{Part96} Particle Data Group, \emph{Review of Particle Physics},
{\it European Phys. J.(C)}, {\bf 15}, (2000) 1-878.

\bibitem{Ruch94} R. Ruchti and M. Wayne, \emph{Quark and Lepton Masses}, 
 Proc. Div. Part. and Fields
APS, S. Seidel, Ed.,  {\it The Albuquerque Meeting} {\bf 2}  (World Scientific, 1994) p. 1220.

\bibitem{Witt89} E. Witten, \emph{Quantum field theory and the Jones polynomial},
{\it Commun. Math. Phys.} {\bf 121} (1989) 351.

\bibitem{Baez94} J. Baez, and P. Muniain, {\em Gauge Fields, Knots
and Gravity\/}  (World Scientific Publishing Co. Sinapore, 1994) p.
322ff.

\bibitem{Kauf91} L. H. Kauffman, {\em Knots and Physics\/} (World
Scientific Publishing Co. Sinapore, 1991).

\bibitem{Sawi96} S. Sawin, {\it Bull. Am. Math. Soc.} {\bf 33} (1996) 413-445.
See p. 428. Also \emph{Links, Quantum Groups and TQFT's}, q-alg/9506002.

\bibitem{Hawk76} S. W. Hawking and G. F. R. Ellis, {\em
The large scale structure of space-time\/} (Cambridge University Press,
London 1976) p. 102ff. 

\bibitem{Mill} R. C. Millikan and D. C. Richman, {\em On the masses of the
leptons, bosons, and quarks}, hep-th/0106106.


\bibitem{Misn73} C. W. Misner, K. S. Thorne, and J. A. Wheeler, {\em
Gravitation\/} (W. H. Freeman and Company, San Francisco 1973) p.
1148ff. 


\bibitem{Adam71} Dale A. Adams, U. S. Pat. No. 3,586,413 \emph{Apparatus
for Providing Energy Communication between a Moving and a Stationary
Terminal} 1971.

\end{thebibliography}
\end{document}